\documentclass[aps,prl,twocolumn]{revtex4}
\usepackage{graphicx}
\usepackage{epstopdf}
\begin{document}
\title{Dynamics of magnetic charges in artificial spin ice}
\author{Paula Mellado}
\author{Olga Petrova}
\author{Yichen Shen}
\author{Oleg Tchernyshyov}
\affiliation{Department of Physics and Astronomy, 
The Johns Hopkins University,
Baltimore, Maryland 21218, USA}

\begin{abstract}
Artificial spin ice has been recently implemented in two-dimensional arrays of mesoscopic magnetic wires.  We propose a theoretical model of magnetization dynamics in artificial spin ice under the action of an applied magnetic field.  Magnetization reversal is mediated by domain walls carrying two units of magnetic charge.  They are emitted by lattice junctions when the the local field exceeds a critical value $H_c$ required to pull apart magnetic charges of opposite sign.  Positive feedback from Coulomb interactions between magnetic charges induces avalanches in magnetization reversal.  
\end{abstract}

\maketitle

Spin ice \cite{Science.294.1495} shares some remarkable properties with 
water ice \cite{petrenko.book}: both possess a 
very large number of low-energy, nearly degenerate configurations satisfying
the Bernal-Fowler ice rules. In water ice, an O$^{2-}$ ion has two protons 
nearby and two farther away; in spin ice, two spins point into and two away 
from the center of every tetrahedron of magnetic ions.  Because the ice rules 
are satisfied by a large fraction of states, the system retains much entropy down to very low temperatures \cite{nature.399.333}.  
Low-frequency dynamics in ice is associated with the motion of defects 
violating the ice rules.  In water ice, these defects carry fractional 
electric charges of $\pm 0.62 e$ (ionic defects) and $\pm 0.38 e$ (Bjerrum defects) \cite{petrenko.book}.  
Fractionalization takes an even more surprising form in spin ice: while
the original degrees of freedom are magnetic \textit{di}poles, the defects 
are magnetic \textit{mono}poles \cite{Nature.451.42, NatPhys.5.258, Science.326.411, Science.326.415, Nature.461.956}.

The charge of an ice defect is defined in terms of the net flux of electric field $\mathbf E$ or magnetic field $\mathbf H$ emerging from the defect.  
On the atomic scale, the flux is obscured by the fields of background ionic 
charges or magnetic dipoles.  Coarse graining is required to reveal the field 
flux of a defect on longer length scales \cite{NatPhys.5.258}.  An alternative 
approach is to alter the model by stretching point-like spin dipoles into 
dumbbell magnets until they touch one another, while keeping their dipole 
moments fixed \cite{Nature.451.42}.  At the expense of a slight change in
the Hamiltonian, the magnetic charge of a defect becomes well defined even 
on the microscopic scale.  It equals $\pm 2q \equiv \pm 2\mu/a$, where $\mu$ 
is the dipole moment and $a$ is the length of a dumbbell.  

\begin{figure}
\includegraphics[width=0.99\columnwidth]{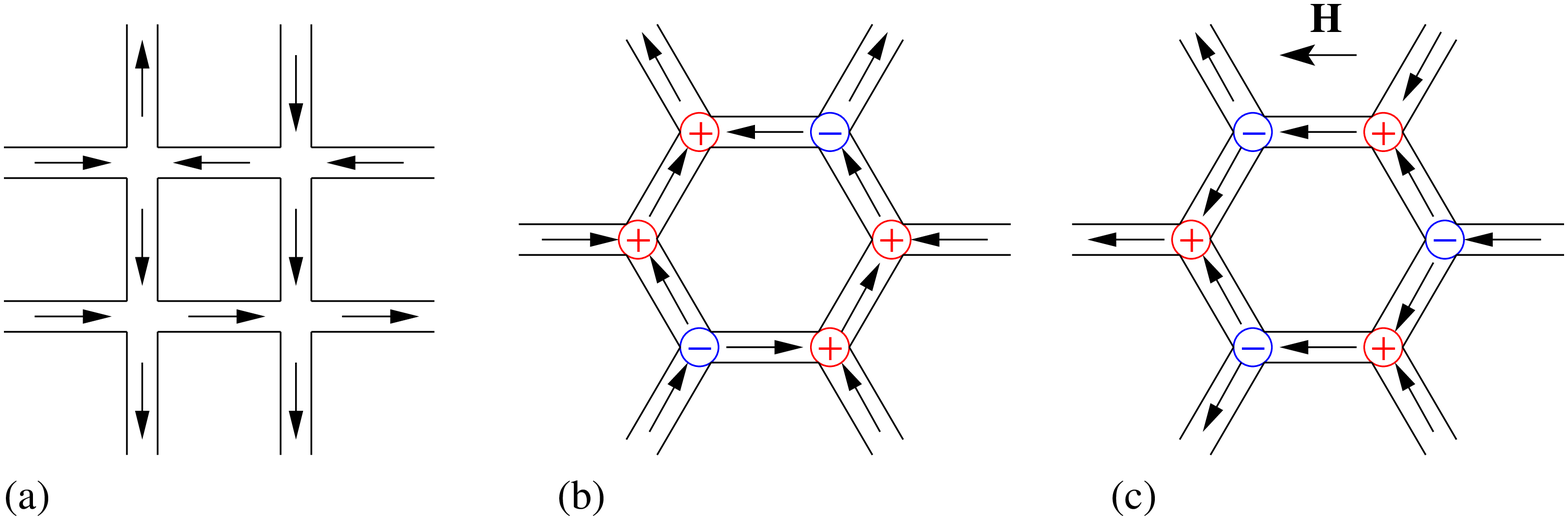}
\caption{(a) A configuration of square spin ice with no magnetic charges.  (b) Honeycomb spin ice always has magnetic charges.  (c) Magnetized honeycomb spin ice.}
\label{fig:network}
\end{figure}

The dumbbell model is realized in artificial spin ice, a network of 
submicron ferromagnetic islands \cite{nature.439.303} or wires 
\cite{tanaka:052411, qi:094418, Ladak:2010}.  Each element represents a 
spin whose 
magnetic dipole moment is aligned with the wire by shape anisotropy, 
Fig.~\ref{fig:network}.  The magnetostatic energy is a positive definite 
quantity $E_\mathrm{dip} = (1/8\pi)\int H^2 dV$, where the integral is 
taken over the entire space.  It is minimized when the magnetic field $\mathbf H = 0$.  $E_\mathrm{dip}$ can be expressed as the Coulomb interaction of magnetic charges with density $\rho(\mathbf r) \equiv \nabla \cdot \mathbf H/4\pi = -\nabla \cdot \mathbf M$.  The field is zero, and the energy is minimized, when there are no magnetic charges.  This yields the ice rule: a network node with zero magnetic charge has zero influx of magnetization.  The zero-flux rule can be satisfied in square ice, Fig.~\ref{fig:network}(a), but not in honeycomb ice, 
Fig.~\ref{fig:network}(b), also known as kagome ice, where the allowed values of magnetic charge $Q$ on a site are $\pm q$ and $\pm 3q$ in units of $q \equiv MA$, where $M$ is the magnetization of the magnetic wire and 
$A$ is its cross section.  Minimization of magnetic charge restricts $Q$
to the values of $\pm q$, yielding the modified ice rule for this lattice:
two arrows in and one out, or vice versa \cite{tanaka:052411, qi:094418}.

The presence of residual magnetic charges in honeycomb ice even at low temperatures may result in a sequence of two phase transitions as its temperature is lowered: magnetic charge order appears first, spin order arises later \cite{PhysRevB.80.140409, arXiv:0906.4781}.  Unfortunately, thermal fluctuations 
are virtually absent in artificial spin ice: reversing the direction of 
magnetization in a single wire requires going over an 
energy barrier of a few million kelvins \cite{nature.439.303}.  Left to
itself, the system remains forever in the same magnetic microstate.  Wang
\textit{et al.} suggested a way to introduce magnetization dynamics into
artificial spin ice by placing the system in a rotating magnetic field of an
oscillating magnitude \cite{PhysRevB.65.092409, ke:037205}, the analog of 
fluidizing granular matter through vibration.  It has been suggested 
\cite{PhysRevLett.98.217203,JApplPhys.106.063913} that such induced dynamics of magnetization effectively create a thermal ensemble with an effective temperature.  

In this Letter we present an entirely different approach to the dynamics of artificial spin ice that incorporates the physics of magnetization reversal in ferromagnetic nanowires, a process mediated by the creation, propagation, and annihilation of magnetic domain walls \cite{ThiavilleBook06, hayashi:2006}.   It is inherently dissipative \cite{tretiakov:127204, PhysRevB.81.060404}: as a domain wall propagates, magnetic energy is transferred to the lattice.  Like fluidized granular matter, artificial spin ice is a system far out of equilibrium and it is not obvious that it can be described in the framework of equilibrium thermodynamics \cite{PhysRevLett.96.237202}.  Mesoscopic degrees of freedom of spin ice tend to move downhill in the energy landscape until they come to rest at a local energy minimum.  We use this approach to describe the dynamics of magnetization observed in honeycomb spin ice \cite{qi:094418} in an applied field.   

In static equilibrium, artificial spin ice is fully described by specifying the direction of the magnetization vector in every link of the lattice.  These are Ising variables because magnetization is aligned with the wire.  Sites of the lattice carry magnetic charge of $\pm q$ or $\pm 3q$ as explained above.  Site charges can be deduced from magnetization variables because the magnetic charge equals the net influx of magnetization.  The converse is not true because the number of links exceeds the number of sites by a factor of 3/2, so the magnetic state of artificial spin ice cannot be described in terms of charges alone \cite{qi:094418}.  Spin variables must be specified for a complete description.

\begin{figure}
\includegraphics[width=0.98\columnwidth]{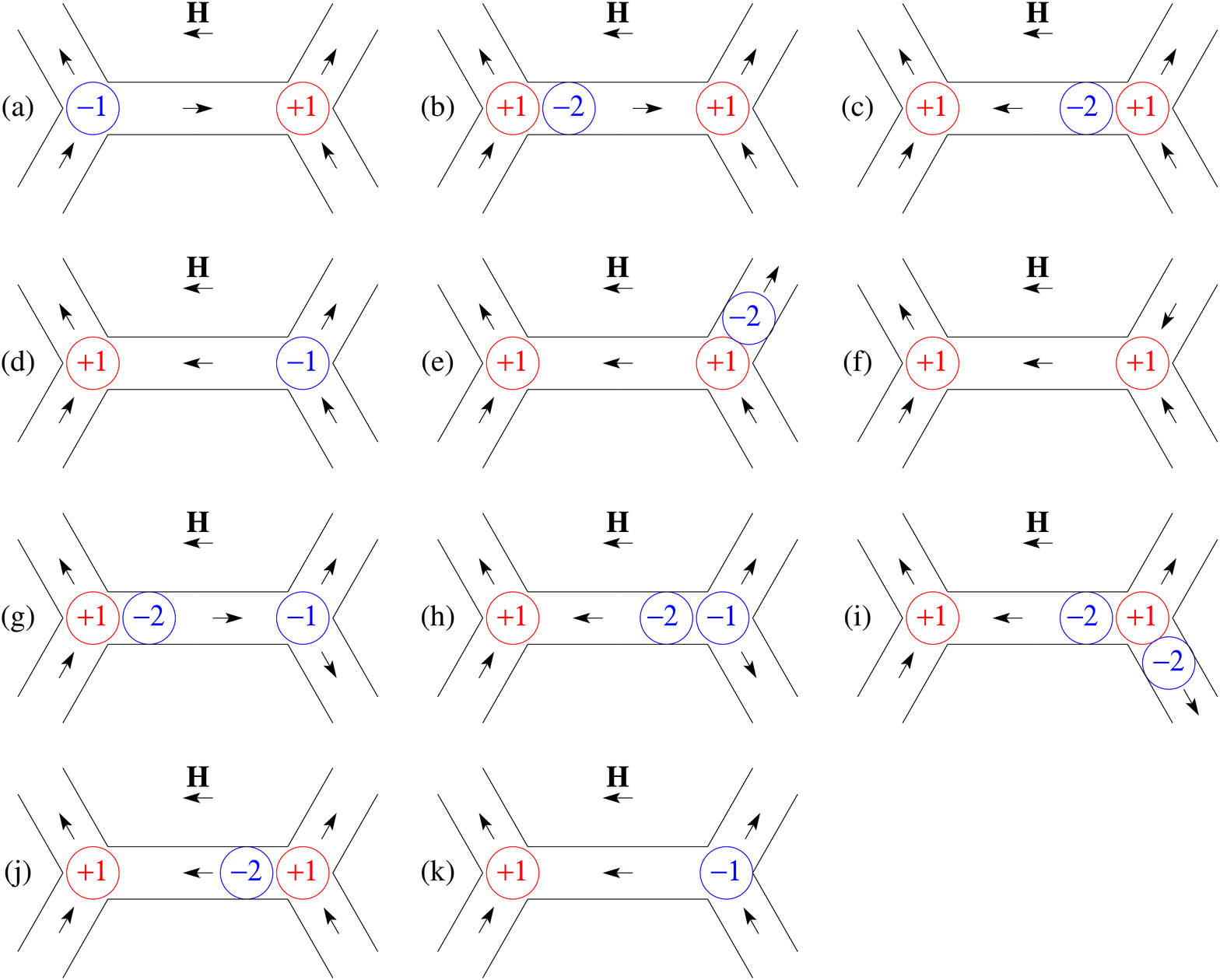}
\caption{Magnetization reversal in honeycomb spin ice.  (a-d) A domain wall is emitted at one end of a link, travels to the other end, and gets absorbed at the junction.  (e-f) If the applied field is sufficiently strong, a new domain wall can be emitted into an adjacent link triggering its magnetization reversal.  (g-k) When a domain wall encounters a site with like magnetic charge, it induces the emission of a new domain wall into an adjacent link.}
\label{fig:link}
\end{figure}

Transitions between static states, triggered by the application of an external magnetic field, involve intermediate states in which the magnetization of one or more links is being reversed.  At the mesoscopic level of our theory, such links are pictured as having two sections uniformly magnetized in opposite directions separated by a domain wall of magnetic charge $Q=\pm 2q$ \cite{tretiakov:127204}.  The reversal of magnetization in a link begins with the creation of a domain wall at one of the link ends.  The process conserves magnetic charge: when a site with magnetic charge $-q$ emits a domain wall of charge $-2q$, the charge of the site changes to $+q$, Fig.~\ref{fig:link}.  The Zeeman force $-2q \mu_0 H$ then pushes the domain wall to the opposite end of the link.

The critical field required to initiate the reversal can be estimated as follows.  A site of charge $+q$ and a domain wall $-2q$ attract each other with a Coulomb force $F \sim \mu_0 2q^2/(4\pi r^2)$ at distances $r$ exceeding the characteristic size of the charges $a$.  The attraction weakens for short distances $r \lesssim a$ when the two charges merge.  The maximum attraction is thus $F_\mathrm{max} \approx \mu_0 2q^2/(4\pi a^2)$.  To pull the charges apart, the Zeeman force $2q B$ from the applied field must exceed $F_\mathrm{max}$,
giving the critical field 
\begin{equation}
H_c = q/(4\pi a^2) = Mtw/(4\pi a^2).
\end{equation}
Domain walls in nanowires of submicron width $w$ have the characteristic size $a \approx 0.6w$ \cite{JApplPhys.99.08G107}.  For the permalloy honeycomb network of Qi \textit{et al.} \cite{qi:094418} with magnetization $M = 8.6 \times 10^5$ A/m, width $w = 110$ nm and thickness $t = 23$ nm, $\mu_0 H_c \approx 50$ mT.  

When the magnetic field is applied at an angle $\theta$ to a link, the Zeeman force comes from the longitudinal component $H \cos{\theta}$.  For this reason we expected the reversal to occur at a higher field $H(\theta) = H_c/\cos{\theta}$.  A similar angular dependence has been observed in magnetic wires with submicron width \cite{PhysRevLett.77.1873}. 

\begin{figure}
\includegraphics[width=0.95\columnwidth]{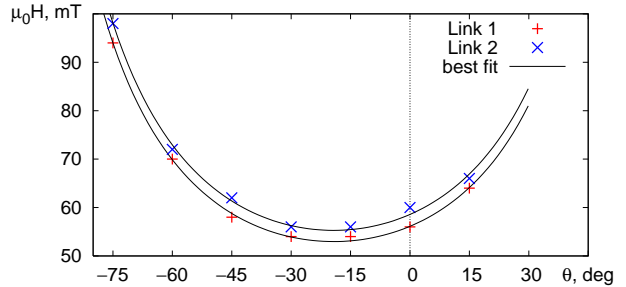}
\caption{
The reversal field $H$ of two out of three magnetic wires forming a junction 
vs. the angle between the field and the axis of the wire whose magnetization is being reversed.  The lines are fits to Eq.~(\ref{eq:Hc-offset}) with $\mu_0 H_c = 52.0$ and 55.3 mT.  
} 
\label{fig:Hc-junction}
\end{figure}

To test this phenomenological model, we performed numerical simulations of magnetization reversal in a single junction of three ferromagnetic nanowires using micromagnetics software package OOMMF \cite{oommf} with the cell size of 2 nm $\times$ 2 nm $\times$ 23 nm.  The dependence $H(\theta)$ is not symmetric, Fig.~\ref{fig:Hc-junction}, and is fit well by the function
\begin{equation}
H(\theta) = H_c/\cos{(\theta + \alpha)},
\label{eq:Hc-offset}
\end{equation}
where the offset $\alpha = 19^\circ$ reflects an asymmetric distribution of magnetization at the junction, as we will discuss elsewhere \cite{mellado.unpublished}.  The critical-field parameter $H_c$ varied slightly between links reflecting small random variations of the width caused by lattice discretization.  Two links of the same junction exhibited slightly different critical fields $H_c$, Fig.~\ref{fig:Hc-junction}.

We use these phenomenological considerations and micromagnetics simulations to build a discrete mesoscopic model of magnetization dynamics in artificial spin ice.  We start with a fully magnetized state in which links of the same orientation have the same direction of magnetization and magnetic charges form a staggered pattern.  Such a state can be obtained by placing the system in a strong magnetic field, Fig.~\ref{fig:network}(c).  In this state, each magnetic wire has uniform magnetization pointing along the wire's axis and each junction contains a magnetic charge of $\pm 1$ in the units of $q=Mtw$ determined by the flux of magnetization into the junction.   The external field is then applied in the opposite direction with a gradually increasing magnitude.  Magnetization reversal begins when the net field $\mathbf H_\mathrm{net}$ at one of the junctions exceeds a critical value determined by Eq.~\ref{eq:Hc-offset}.  The net magnetic field $\mathbf H_\mathrm{net}$ is a superposition of the applied field $\mathbf H_\mathrm{app}$ and of the demagnetizing field of the sample $\mathbf H_\mathrm{dem}$.  The latter is computed as a sum of Coulombic fields of individual junctions, $\mathbf H = Q\mathbf r/(4\pi r^3)$. The junction, initially containing charge $\pm 1$, emits a domain wall with charge $\pm 2$ and changes its own charge to $\mp 1$.  The emitted domain wall is pushed by the magnetic field to the other end of the link, reversing the link magnetization in the process, Fig.~\ref{fig:link} (b-c).  Quenched disorder, inevitably present in real samples, is modeled by setting at random slightly different critical fields $H_c$ in individual wires with a mean $\bar H_c$ and a distribution width $\Delta H_c$.  

As the domain wall with charge $\pm 2$ reaches the other end of the link, its further fate depends on sign of the magnetic charge it meets at the junction.  If the charge is of opposite sign, $\mp 1$, then the domain wall is absorbed by the junction, Fig.~\ref{fig:link}(c-d), whose charge reverts to $\pm 1$.  If the net field is strong enough to stimulate the emission of a new domain wall of charge $\pm 2$ out of this junction, Fig.~\ref{fig:link}(e), one of the adjacent links reverses its magnetization, Fig.~\ref{fig:link}(f).  Otherwise, the evolution stops at the stage shown in Fig.~\ref{fig:link}(d).

\begin{figure}
\includegraphics[width=0.98\columnwidth]{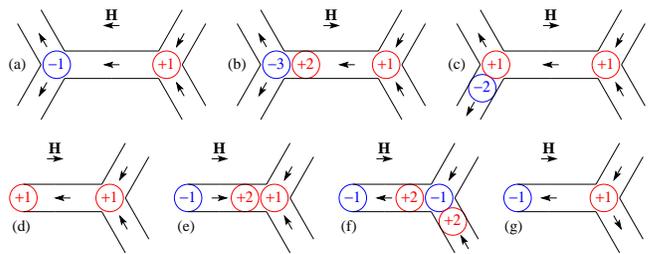}
\caption{Magnetization reversal in uniformly magnetized spin ice.  (a-b) In the bulk, the reversal in a link magnetized against the field would lead to the formation of triple charge, which can only happen when the field is of order $3 H_c$.  (c) Instead, the reversal occurs first in links magnetized at $120^\circ$ to the field when $H \approx 2 H_c$.  (d-g) At the edge, the reversal begins when $H \approx H_c$ and propagates into the bulk.}
\label{fig:link2}
\end{figure}

Alternatively, if the domain wall comes to a junction with the same sign of charge, Fig.~\ref{fig:link2}(a-b), it stops distance $a$ short of the junction thanks to magnetostatic repulsion.  While this could be a new equilibrium position, the charged domain wall creates a field of strength $2H_c$ at the junction, so that the net field at the junction is close to $3H_c$.  Its projection onto an adjacent link, $1.5 H_c$, is sufficient to stimulate the emission of a new domain wall of charge $\pm 2$ into that link, Fig.~\ref{fig:link2}(c).  The junction, now carrying charge of the opposite sign, $\mp 1$, pulls in the original domain wall and settles down in a state with charge $\pm 1$, Fig.~\ref{fig:link2}(d).  

The sequence illustrated in Fig.~\ref{fig:link2} explains why ice rule violations are hard to find in honeycomb ice of Qi \textit{et al.} \cite{qi:094418}.  Unless variations of the critical field are so strong that $H_c$ at some junctions exceeds $1.5\bar H_c$, triply charged junctions, Fig.~\ref{fig:link2}(b), are unstable and decay via the emission of a new domain wall, Fig.~\ref{fig:link2}(c-d).  Permalloy samples of Qi \textit{et al.} exhibit a Gaussian distribution of critical fields with a standard deviation $\Delta H_c = 0.04 \bar H_c$ \cite{daunheimer.unpublished}, so that states with charge $\pm 3$ are only transients.  Much stronger disorder exists in cobalt samples of Ladak \textit{et al.} \cite{Ladak:2010} who observed magnetization reversal in a field range between $H = 50$ and 75 mT.  Thus some of the domain walls encounter junctions whose critical field exceeds $1.5 H$, which explains the presence of charges $\pm 3$.

In the limit of weak disorder, $\Delta H_c \ll \bar H_c$, there is another characteristic scale of the field that becomes important. The new scale set by the demagnetizing field of the sample $\mathbf H_\mathrm{dem}$, is the strength of the field created by a unit magnetic charge, $Q = Mtw$, at a neighboring junction distance $L$ away, $H_0 = Mtw/(4\pi L^2)$.  When $\Delta H_c  \gg H_0$, the reversal of magnetization is controlled mostly by the effects of quenched disorder, with links reversing in a largely independent fashion in the order of increasing critical field $H_c$.  Conversely, when $\Delta H_c \ll H_0$, the reversal proceeds in a correlated fashion because of a positive feedback: the reversal of magnetization in one link redistributes magnetic charges at its ends, which in turn increases the net field at adjacent junctions and thus triggers the emission of domain walls there.  In samples of Qi \textit{et al.}, $H_0 = 0.87$ mT, which is comparable to the width of their reversal region, $\Delta H_c = 2$ mT.  

\begin{figure}
\includegraphics[width=0.95\columnwidth]{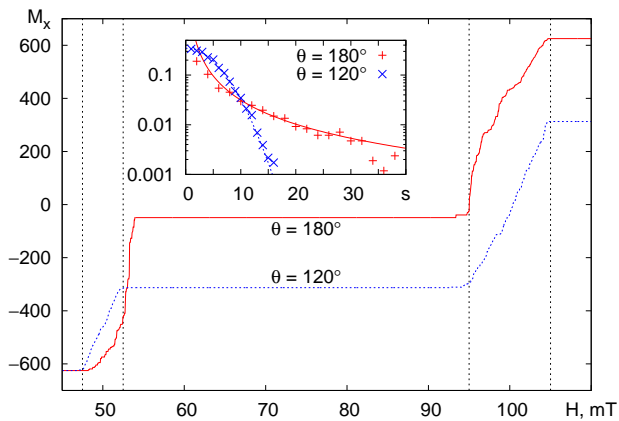}
\caption{Simulated magnetization reversals.  A sample is initially magnetized in a strong field directed as in Fig.~\ref{fig:network}(c).  Subsequently, the field is switched off and reapplied at angles $\theta = 120^\circ$ and $180^\circ$ to the initial direction.  Vertical dashed lines are at $\bar H_c \pm \Delta H_c/2$ and $2\bar H_c \pm \Delta H_c$.  Inset: Distribution of avalanche lengths $D(s)$ in the range of fields near $\bar H_c$.  Deviations near the bottom of the graph are due to statistical noise.}
\label{fig:hysteresis}
\end{figure}

We simulated magnetization reversal in this model with the critical fields uniformly distributed in an interval of width $\Delta H_c = 5$ mT around the mean $\bar H_c = 50$ mT and the Coulomb field scale $H_0 = 0.87$ mT.  For simplicity we set the offset angle $\alpha = 0$.  A sample containing 937 links was initially magnetized along one subset of links, Fig.~\ref{fig:network}(c).  Subsequently, the field was switched off and a reversal curve $\mathbf M(H)$ was measured in field rotated through angle $\theta$ from the initial direction.  For $120^\circ$, quenched disorder dominates so that magnetization reversals occur largely independently, in two stages.  Links magnetized against the field switch when the applied field is within the range $\bar H_c \pm \Delta H_c/2$, whereas links magnetized at $120^\circ$ to the field switch in the range $2\bar H_c \pm \Delta H_c$.  The net magnetization $M_x$ grows in an approximately linear fashion in both ranges, Fig.~\ref{fig:hysteresis}, as expected for links with a uniform distribution of $H_c$.  Links do not reverse completely independently from one another: as noted previously, the redistribution of magnetic charges induced by the reversal of magnetization in one link may trigger another reversal nearby.  We observed that reversals often involves small groups of links.   As can be seen in the inset of Fig.~\ref{fig:hysteresis}, the distribution of the number of links $s$ reversing in a single event is Gaussian, $D(s) \propto \exp{(-s^2/2\xi^2)}$, with $\xi = 4.6$.  

An entirely different process is observed when the field is rotated through $\theta = 180^\circ$.  In this case, Coulomb interactions play a major role and the reversal proceeds through avalanches evidenced by steps in $M_x(H)$, Fig.~\ref{fig:hysteresis}.  When the field is near $\bar H_c$, the reversal cannot begin in the bulk because links parallel to the applied field have the wrong sign of magnetic charges at the ends and will reverse only in a much higher field (of order $3H_c$).  Links at the edges have no such problem and the reversal starts when a site at the edge emits a domain wall, Fig.~\ref{fig:link2}(d-e).  When the domain wall reaches the other end of the link, it encounters a site with like magnetic charge and triggers the emission of a new domain wall, Fig.~\ref{fig:link2}(f), and the reversal of magnetization in an adjacent link, Fig.~\ref{fig:link2}(g).  This triggers an avalanche of reversals that stops when the traveling domain wall is absorbed by a junction with a large critical field $H_c$ or runs into already reversed links \cite{Ladak:2010,NatPhys.6.323}.  The distribution of avalanche lengths (Fig.~\ref{fig:hysteresis}) fits a power law, $D(s) \propto s^{-\tau}$, with the exponent $\tau = 1.6$, indicative of self-organized criticality \cite{PhysRevA.38.364}.  Chain reversals involving 3 links have been reported by Ladak \textit{et al.} \cite{Ladak:2010} in this geometry; avalanches involving up to 39 links have been observed by Daunheimer \textit{et al.} \cite{daunheimer.unpublished}.

We have presented a discrete model of artificial spin ice where magnetization dynamics is mediated by domain walls carrying magnetic charge.  Interactions between magnetic charges compete with the effects of quenched disorder.  In samples with low disorder, positive feedback from charge redistribution is responsible for magnetic avalanches that have been observed in some experimental situations.   

We thank John Cumings for numerous discussions and for sharing his unpublished data.  This work was supported in part by the NSF Grant No. DMR-0520491.

\bibliography{spinice,micromagnetics}

\begin{thebibliography}{30}
\expandafter\ifx\csname natexlab\endcsname\relax\def\natexlab#1{#1}\fi
\expandafter\ifx\csname bibnamefont\endcsname\relax
  \def\bibnamefont#1{#1}\fi
\expandafter\ifx\csname bibfnamefont\endcsname\relax
  \def\bibfnamefont#1{#1}\fi
\expandafter\ifx\csname citenamefont\endcsname\relax
  \def\citenamefont#1{#1}\fi
\expandafter\ifx\csname url\endcsname\relax
  \def\url#1{\texttt{#1}}\fi
\expandafter\ifx\csname urlprefix\endcsname\relax\def\urlprefix{URL }\fi
\providecommand{\bibinfo}[2]{#2}
\providecommand{\eprint}[2][]{\url{#2}}

\bibitem[{\citenamefont{Bramwell and Gingras}(2001)}]{Science.294.1495}
\bibinfo{author}{\bibfnamefont{S.~T.} \bibnamefont{Bramwell}} \bibnamefont{and}
  \bibinfo{author}{\bibfnamefont{M.~J.~P.} \bibnamefont{Gingras}},
  \bibinfo{journal}{Science} \textbf{\bibinfo{volume}{294}},
  \bibinfo{pages}{1495} (\bibinfo{year}{2001}).

\bibitem[{\citenamefont{Petrenko and Whitworth}(1999)}]{petrenko.book}
\bibinfo{author}{\bibfnamefont{V.~F.} \bibnamefont{Petrenko}} \bibnamefont{and}
  \bibinfo{author}{\bibfnamefont{R.~W.} \bibnamefont{Whitworth}},
  \emph{\bibinfo{title}{Physics of ice}} (\bibinfo{publisher}{Oxford University
  Press}, \bibinfo{year}{1999}).

\bibitem[{\citenamefont{Ramirez et~al.}(1999)\citenamefont{Ramirez, Hayashi,
  Cava, Siddharthan, and Shastry}}]{nature.399.333}
\bibinfo{author}{\bibfnamefont{A.~P.} \bibnamefont{Ramirez}},
  \bibinfo{author}{\bibfnamefont{A.}~\bibnamefont{Hayashi}},
  \bibinfo{author}{\bibfnamefont{R.~J.} \bibnamefont{Cava}},
  \bibinfo{author}{\bibfnamefont{R.}~\bibnamefont{Siddharthan}},
  \bibnamefont{and} \bibinfo{author}{\bibfnamefont{B.~S.}
  \bibnamefont{Shastry}}, \bibinfo{journal}{Nature}
  \textbf{\bibinfo{volume}{399}}, \bibinfo{pages}{333} (\bibinfo{year}{1999}).

\bibitem[{\citenamefont{Castelnovo et~al.}(2008)\citenamefont{Castelnovo,
  Moessner, and Sondhi}}]{Nature.451.42}
\bibinfo{author}{\bibfnamefont{C.}~\bibnamefont{Castelnovo}},
  \bibinfo{author}{\bibfnamefont{R.}~\bibnamefont{Moessner}}, \bibnamefont{and}
  \bibinfo{author}{\bibfnamefont{S.~L.} \bibnamefont{Sondhi}},
  \bibinfo{journal}{Nature} \textbf{\bibinfo{volume}{451}}, \bibinfo{pages}{42}
  (\bibinfo{year}{2008}).

\bibitem[{\citenamefont{Jaubert and Holdsworth}(2009)}]{NatPhys.5.258}
\bibinfo{author}{\bibfnamefont{L.~D.~C.} \bibnamefont{Jaubert}}
  \bibnamefont{and} \bibinfo{author}{\bibfnamefont{P.~C.~W.}
  \bibnamefont{Holdsworth}}, \bibinfo{journal}{Nat. Phys.}
  \textbf{\bibinfo{volume}{5}}, \bibinfo{pages}{258} (\bibinfo{year}{2009}).

\bibitem[{\citenamefont{{D. Morris \textit{et al.}}}(2009)}]{Science.326.411}
\bibinfo{author}{\bibnamefont{{D. Morris \textit{et al.}}}},
  \bibinfo{journal}{Science} \textbf{\bibinfo{volume}{326}},
  \bibinfo{pages}{411} (\bibinfo{year}{2009}).

\bibitem[{\citenamefont{Fennell et~al.}(2009)\citenamefont{Fennell, Deen,
  Wildes, Schmalzl, Prabhakaran, Boothroyd, Aldus, Mcmorrow, and
  Bramwell}}]{Science.326.415}
\bibinfo{author}{\bibfnamefont{T.}~\bibnamefont{Fennell}},
  \bibinfo{author}{\bibfnamefont{P.}~\bibnamefont{Deen}},
  \bibinfo{author}{\bibfnamefont{A.}~\bibnamefont{Wildes}},
  \bibinfo{author}{\bibfnamefont{K.}~\bibnamefont{Schmalzl}},
  \bibinfo{author}{\bibfnamefont{D.}~\bibnamefont{Prabhakaran}},
  \bibinfo{author}{\bibfnamefont{A.}~\bibnamefont{Boothroyd}},
  \bibinfo{author}{\bibfnamefont{R.}~\bibnamefont{Aldus}},
  \bibinfo{author}{\bibfnamefont{D.}~\bibnamefont{Mcmorrow}}, \bibnamefont{and}
  \bibinfo{author}{\bibfnamefont{S.}~\bibnamefont{Bramwell}},
  \bibinfo{journal}{Science} \textbf{\bibinfo{volume}{326}},
  \bibinfo{pages}{415} (\bibinfo{year}{2009}).

\bibitem[{\citenamefont{Bramwell et~al.}(2009)\citenamefont{Bramwell, Giblin,
  Calder, Aldus, Prabhakaran, and Fennell}}]{Nature.461.956}
\bibinfo{author}{\bibfnamefont{S.~T.} \bibnamefont{Bramwell}},
  \bibinfo{author}{\bibfnamefont{S.~R.} \bibnamefont{Giblin}},
  \bibinfo{author}{\bibfnamefont{S.}~\bibnamefont{Calder}},
  \bibinfo{author}{\bibfnamefont{R.}~\bibnamefont{Aldus}},
  \bibinfo{author}{\bibfnamefont{D.}~\bibnamefont{Prabhakaran}},
  \bibnamefont{and} \bibinfo{author}{\bibfnamefont{T.}~\bibnamefont{Fennell}},
  \bibinfo{journal}{Nature} \textbf{\bibinfo{volume}{456}},
  \bibinfo{pages}{956} (\bibinfo{year}{2009}).

\bibitem[{\citenamefont{{R. F. Wang \textit{et al.}}}(2006)}]{nature.439.303}
\bibinfo{author}{\bibnamefont{{R. F. Wang \textit{et al.}}}},
  \bibinfo{journal}{Nature} \textbf{\bibinfo{volume}{439}},
  \bibinfo{pages}{303} (\bibinfo{year}{2006}).

\bibitem[{\citenamefont{Tanaka et~al.}(2006)\citenamefont{Tanaka, Saitoh,
  Miyajima, Yamaoka, and Iye}}]{tanaka:052411}
\bibinfo{author}{\bibfnamefont{M.}~\bibnamefont{Tanaka}},
  \bibinfo{author}{\bibfnamefont{E.}~\bibnamefont{Saitoh}},
  \bibinfo{author}{\bibfnamefont{H.}~\bibnamefont{Miyajima}},
  \bibinfo{author}{\bibfnamefont{T.}~\bibnamefont{Yamaoka}}, \bibnamefont{and}
  \bibinfo{author}{\bibfnamefont{Y.}~\bibnamefont{Iye}},
  \bibinfo{journal}{Phys. Rev. B} \textbf{\bibinfo{volume}{73}},
  \bibinfo{pages}{052411} (\bibinfo{year}{2006}).

\bibitem[{\citenamefont{Qi et~al.}(2008)\citenamefont{Qi, Brintlinger, and
  Cumings}}]{qi:094418}
\bibinfo{author}{\bibfnamefont{Y.}~\bibnamefont{Qi}},
  \bibinfo{author}{\bibfnamefont{T.}~\bibnamefont{Brintlinger}},
  \bibnamefont{and} \bibinfo{author}{\bibfnamefont{J.}~\bibnamefont{Cumings}},
  \bibinfo{journal}{Phys. Rev. B} \textbf{\bibinfo{volume}{77}},
  \bibinfo{pages}{094418} (\bibinfo{year}{2008}).

\bibitem[{\citenamefont{Ladak et~al.}(2010)\citenamefont{Ladak, Read, Perkins,
  Cohen, and Branford}}]{Ladak:2010}
\bibinfo{author}{\bibfnamefont{S.}~\bibnamefont{Ladak}},
  \bibinfo{author}{\bibfnamefont{D.~E.} \bibnamefont{Read}},
  \bibinfo{author}{\bibfnamefont{G.~K.} \bibnamefont{Perkins}},
  \bibinfo{author}{\bibfnamefont{L.~F.} \bibnamefont{Cohen}}, \bibnamefont{and}
  \bibinfo{author}{\bibfnamefont{W.~R.} \bibnamefont{Branford}},
  \bibinfo{journal}{Nat. Phys.} \textbf{\bibinfo{volume}{6}},
  \bibinfo{pages}{359} (\bibinfo{year}{2010}).

\bibitem[{\citenamefont{M{\"o}ller and Moessner}(2009)}]{PhysRevB.80.140409}
\bibinfo{author}{\bibfnamefont{G.}~\bibnamefont{M{\"o}ller}} \bibnamefont{and}
  \bibinfo{author}{\bibfnamefont{R.}~\bibnamefont{Moessner}},
  \bibinfo{journal}{Phys. Rev. B} \textbf{\bibinfo{volume}{80}},
  \bibinfo{pages}{140409} (\bibinfo{year}{2009}).

\bibitem[{\citenamefont{Chern et~al.}(unpublished)\citenamefont{Chern, Mellado,
  and Tchernyshyov}}]{arXiv:0906.4781}
\bibinfo{author}{\bibfnamefont{G.-W.} \bibnamefont{Chern}},
  \bibinfo{author}{\bibfnamefont{P.}~\bibnamefont{Mellado}}, \bibnamefont{and}
  \bibinfo{author}{\bibfnamefont{O.}~\bibnamefont{Tchernyshyov}}
  (\bibinfo{year}{unpublished}), \eprint{arXiv:0906.4781}.

\bibitem[{\citenamefont{Cowburn}(2002)}]{PhysRevB.65.092409}
\bibinfo{author}{\bibfnamefont{R.~P.} \bibnamefont{Cowburn}},
  \bibinfo{journal}{Phys. Rev. B} \textbf{\bibinfo{volume}{65}},
  \bibinfo{pages}{092409} (\bibinfo{year}{2002}).

\bibitem[{\citenamefont{Ke et~al.}(2008)\citenamefont{Ke, Li, Nisoli, Lammert,
  McConville, Wang, Crespi, and Schiffer}}]{ke:037205}
\bibinfo{author}{\bibfnamefont{X.}~\bibnamefont{Ke}},
  \bibinfo{author}{\bibfnamefont{J.}~\bibnamefont{Li}},
  \bibinfo{author}{\bibfnamefont{C.}~\bibnamefont{Nisoli}},
  \bibinfo{author}{\bibfnamefont{P.~E.} \bibnamefont{Lammert}},
  \bibinfo{author}{\bibfnamefont{W.}~\bibnamefont{McConville}},
  \bibinfo{author}{\bibfnamefont{R.~F.} \bibnamefont{Wang}},
  \bibinfo{author}{\bibfnamefont{V.~H.} \bibnamefont{Crespi}},
  \bibnamefont{and} \bibinfo{author}{\bibfnamefont{P.}~\bibnamefont{Schiffer}},
  \bibinfo{journal}{Phys. Rev. Lett.} \textbf{\bibinfo{volume}{101}},
  \bibinfo{pages}{037205} (\bibinfo{year}{2008}).

\bibitem[{\citenamefont{Nisoli et~al.}(2007)\citenamefont{Nisoli, Wang, Li,
  McConville, Lammert, Schiffer, and Crespi}}]{PhysRevLett.98.217203}
\bibinfo{author}{\bibfnamefont{C.}~\bibnamefont{Nisoli}},
  \bibinfo{author}{\bibfnamefont{R.}~\bibnamefont{Wang}},
  \bibinfo{author}{\bibfnamefont{J.}~\bibnamefont{Li}},
  \bibinfo{author}{\bibfnamefont{W.~F.} \bibnamefont{McConville}},
  \bibinfo{author}{\bibfnamefont{P.~E.} \bibnamefont{Lammert}},
  \bibinfo{author}{\bibfnamefont{P.}~\bibnamefont{Schiffer}}, \bibnamefont{and}
  \bibinfo{author}{\bibfnamefont{V.~H.} \bibnamefont{Crespi}},
  \bibinfo{journal}{Phys. Rev. Lett.} \textbf{\bibinfo{volume}{98}},
  \bibinfo{pages}{217203} (\bibinfo{year}{2007}).

\bibitem[{\citenamefont{M{\'o}l et~al.}(2009)\citenamefont{M{\'o}l, Silva,
  Silva, Pereira, Moura-Melo, and Costa}}]{JApplPhys.106.063913}
\bibinfo{author}{\bibfnamefont{L.~A.} \bibnamefont{M{\'o}l}},
  \bibinfo{author}{\bibfnamefont{R.~L.} \bibnamefont{Silva}},
  \bibinfo{author}{\bibfnamefont{R.~C.} \bibnamefont{Silva}},
  \bibinfo{author}{\bibfnamefont{A.~R.} \bibnamefont{Pereira}},
  \bibinfo{author}{\bibfnamefont{W.~A.} \bibnamefont{Moura-Melo}},
  \bibnamefont{and} \bibinfo{author}{\bibfnamefont{B.~V.} \bibnamefont{Costa}},
  \bibinfo{journal}{J. Appl. Phys.} \textbf{\bibinfo{volume}{106}},
  \bibinfo{pages}{063913} (\bibinfo{year}{2009}).

\bibitem[{\citenamefont{Thiaville and Nakatani}(2006)}]{ThiavilleBook06}
\bibinfo{author}{\bibfnamefont{A.}~\bibnamefont{Thiaville}} \bibnamefont{and}
  \bibinfo{author}{\bibfnamefont{Y.}~\bibnamefont{Nakatani}}, in
  \emph{\bibinfo{booktitle}{Spin Dynamics in Confined Magnetic Structures III}}
  (\bibinfo{publisher}{Springer}, \bibinfo{year}{2006}), vol.
  \bibinfo{volume}{101} of \emph{\bibinfo{series}{Topics in applied physics}},
  pp. \bibinfo{pages}{161--205}.

\bibitem[{\citenamefont{Hayashi et~al.}(2007)\citenamefont{Hayashi, Thomas,
  Rettner, Moriya, and Parkin}}]{hayashi:2006}
\bibinfo{author}{\bibfnamefont{M.}~\bibnamefont{Hayashi}},
  \bibinfo{author}{\bibfnamefont{L.}~\bibnamefont{Thomas}},
  \bibinfo{author}{\bibfnamefont{C.}~\bibnamefont{Rettner}},
  \bibinfo{author}{\bibfnamefont{R.}~\bibnamefont{Moriya}}, \bibnamefont{and}
  \bibinfo{author}{\bibfnamefont{S.~S.~P.} \bibnamefont{Parkin}},
  \bibinfo{journal}{Nat. Phys.} \textbf{\bibinfo{volume}{3}},
  \bibinfo{pages}{21} (\bibinfo{year}{2007}).

\bibitem[{\citenamefont{Tretiakov et~al.}(2008)\citenamefont{Tretiakov, Clarke,
  Chern, Bazaliy, and Tchernyshyov}}]{tretiakov:127204}
\bibinfo{author}{\bibfnamefont{O.~A.} \bibnamefont{Tretiakov}},
  \bibinfo{author}{\bibfnamefont{D.}~\bibnamefont{Clarke}},
  \bibinfo{author}{\bibfnamefont{G.-W.} \bibnamefont{Chern}},
  \bibinfo{author}{\bibfnamefont{Y.~B.} \bibnamefont{Bazaliy}},
  \bibnamefont{and}
  \bibinfo{author}{\bibfnamefont{O.}~\bibnamefont{Tchernyshyov}},
  \bibinfo{journal}{Phys. Rev. Lett.} \textbf{\bibinfo{volume}{100}},
  \bibinfo{eid}{127204} (\bibinfo{year}{2008}).

\bibitem[{\citenamefont{Wong and Tserkovnyak}(2010)}]{PhysRevB.81.060404}
\bibinfo{author}{\bibfnamefont{C.~H.} \bibnamefont{Wong}} \bibnamefont{and}
  \bibinfo{author}{\bibfnamefont{Y.}~\bibnamefont{Tserkovnyak}},
  \bibinfo{journal}{Phys. Rev. B} \textbf{\bibinfo{volume}{81}},
  \bibinfo{pages}{060404} (\bibinfo{year}{2010}).

\bibitem[{\citenamefont{M{\"o}ller and Moessner}(2006)}]{PhysRevLett.96.237202}
\bibinfo{author}{\bibfnamefont{G.}~\bibnamefont{M{\"o}ller}} \bibnamefont{and}
  \bibinfo{author}{\bibfnamefont{R.}~\bibnamefont{Moessner}},
  \bibinfo{journal}{Phys. Rev. Lett.} \textbf{\bibinfo{volume}{96}},
  \bibinfo{pages}{237202} (\bibinfo{year}{2006}).

\bibitem[{\citenamefont{Kunz}(2006)}]{JApplPhys.99.08G107}
\bibinfo{author}{\bibfnamefont{A.}~\bibnamefont{Kunz}}, \bibinfo{journal}{J.
  Appl. Phys.} \textbf{\bibinfo{volume}{99}}, \bibinfo{pages}{08G107}
  (\bibinfo{year}{2006}).

\bibitem[{\citenamefont{Wernsdorfer et~al.}(1996)\citenamefont{Wernsdorfer,
  Doudin, Mailly, Hasselbach, Benoit, Meier, Ansermet, and
  Barbara}}]{PhysRevLett.77.1873}
\bibinfo{author}{\bibfnamefont{W.}~\bibnamefont{Wernsdorfer}},
  \bibinfo{author}{\bibfnamefont{B.}~\bibnamefont{Doudin}},
  \bibinfo{author}{\bibfnamefont{D.}~\bibnamefont{Mailly}},
  \bibinfo{author}{\bibfnamefont{K.}~\bibnamefont{Hasselbach}},
  \bibinfo{author}{\bibfnamefont{A.}~\bibnamefont{Benoit}},
  \bibinfo{author}{\bibfnamefont{J.}~\bibnamefont{Meier}},
  \bibinfo{author}{\bibfnamefont{J.~P.} \bibnamefont{Ansermet}},
  \bibnamefont{and} \bibinfo{author}{\bibfnamefont{B.}~\bibnamefont{Barbara}},
  \bibinfo{journal}{Phys. Rev. Lett.} \textbf{\bibinfo{volume}{77}},
  \bibinfo{pages}{1873} (\bibinfo{year}{1996}).

\bibitem[{\citenamefont{Donahue and Porter}(1999)}]{oommf}
\bibinfo{author}{\bibfnamefont{M.~J.} \bibnamefont{Donahue}} \bibnamefont{and}
  \bibinfo{author}{\bibfnamefont{D.~G.} \bibnamefont{Porter}},
  \bibinfo{type}{Tech. Rep.} \bibinfo{number}{NISTIR 6376},
  \bibinfo{institution}{National Institute of Standards and Technology},
  \bibinfo{address}{Gaithersburg, MD} (\bibinfo{year}{1999}),
  \bibinfo{note}{http://math.nist.gov/oommf}.

\bibitem[{\citenamefont{{P. Mellado \textit{et al.}}}()}]{mellado.unpublished}
\bibinfo{author}{\bibnamefont{{P. Mellado \textit{et al.}}}},
  \bibinfo{howpublished}{manuscript in preparation}.

\bibitem[{\citenamefont{Daunheimer
  et~al.}(unpublished)\citenamefont{Daunheimer, Qi, and
  Cumings}}]{daunheimer.unpublished}
\bibinfo{author}{\bibfnamefont{S.}~\bibnamefont{Daunheimer}},
  \bibinfo{author}{\bibfnamefont{Y.}~\bibnamefont{Qi}}, \bibnamefont{and}
  \bibinfo{author}{\bibfnamefont{J.}~\bibnamefont{Cumings}}
  (\bibinfo{year}{unpublished}).

\bibitem[{\citenamefont{Tchernyshyov}(2006)}]{NatPhys.6.323}
\bibinfo{author}{\bibfnamefont{O.}~\bibnamefont{Tchernyshyov}},
  \bibinfo{journal}{Nat. Phys.} \textbf{\bibinfo{volume}{6}},
  \bibinfo{pages}{323} (\bibinfo{year}{2006}).

\bibitem[{\citenamefont{Bak et~al.}(1988)\citenamefont{Bak, Tang, and
  Wiesenfeld}}]{PhysRevA.38.364}
\bibinfo{author}{\bibfnamefont{P.}~\bibnamefont{Bak}},
  \bibinfo{author}{\bibfnamefont{C.}~\bibnamefont{Tang}}, \bibnamefont{and}
  \bibinfo{author}{\bibfnamefont{K.}~\bibnamefont{Wiesenfeld}},
  \bibinfo{journal}{Phys. Rev. A} \textbf{\bibinfo{volume}{38}},
  \bibinfo{pages}{364} (\bibinfo{year}{1988}).

\end{thebibliography}

\end{document}